# Torus-Projected Electromagnetic Wormholes Enabled by Anisotropic Singularity Reconstruction of Metamaterials

Junke Liao[1]†, Wen Xiao[1]†, Yixiao Ge[1], Huanyang Chen[1]*

[1]Department of Physics, Xiamen University; Xiamen 365100, China.

†These authors contributed equally to this work

*Corresponding author. Email: kenyon@xmu.edu.cn

**Abstract:** Transformation optics uses coordinate mappings to emulate curved geometries and control electromagnetic fields. However, existing approaches primarily focus on geometric deformation while offering limited control over the global topology of the resulting optical space. Here we introduce a torus projection combined with a conformal mapping to construct a wormhole-like virtual optical geometry, providing a controllable route to manipulate virtual space topology within transformation optics. By tuning a single torus parameter, the virtual-space refractive index switches between isotropic and anisotropic forms, driving a transition from a disconnected horn-torus geometry to a connected ring-torus geometry in virtual space. In physical space, this virtual-space topology transition gives rise to switchable black-hole-like trapping, field redistribution, and geometry-controlled wave responses. Our results demonstrate how refractive-index anisotropy governs singularity structure and topology in transformation optics, while also suggesting practical functionalities including beam splitting, absorbing-like switching, and curvature-assisted focusing.

## Introduction

Transformation optics (TO) [*1,2*] provides a unified framework linking electromagnetic field control to spacetime geometry through coordinate transformations [*4-7*]. This approach enables the design of materials that mimic gravitational effects, giving rise to analogs of phenomena such as black holes [*8,9*] and gravitational lenses [*10*]. Beyond reproducing known astrophysical metrics, TO offers a route to emulate theoretical spacetime constructs predicted by general relativity. Among these, the Einstein–Rosen bridge, or wormhole, represents a geometry connecting two black holes through a nontrivial topology [*11,12*]. TO-based analogs of such "wormhole-like" systems have been demonstrated as invisible electromagnetic tunnels [*13*] and embedded two-dimensional surfaces [*14,15*], capturing essential geometric and field properties.

More broadly, conformal transformation optics [*16,17*] has provided a useful route to optical-space compactification and singularity engineering [*18,19*], enabling nontrivial effective optical geometries and topological optical responses [*20-22*]. Such geometries are not only of fundamental interest, but have also been considered in relation to practical wave-control functionalities, including cavity mode engineering [*23-25*], wave focusing [*26*] and cloaking [*22,27*], and geometry-assisted control of light propagation [*26,28*]. These developments naturally suggest exploring optical architectures in which topology and singularity structure can be tuned in a controlled manner while retaining observable electromagnetic consequences in physical space.

In this work, we construct a ring-shaped electromagnetic wormhole in virtual space by combining a conformal mapping hosting dual black-hole–like singularities with a torus projection. This construction allows the topology of the virtual optical space to be continuously tuned while the corresponding singularities remain spatially separated in physical space. As the torus parameters are varied, the refractive-index distribution in virtual space switches between isotropic and anisotropic forms, leading to a topological transition of the toroidal geometry and the emergence of geodesic connectivity in virtual space. We show that anisotropy in transformation optics reconstructs optical singularities in virtual space, while its physical-space manifestation appears as switchable black-hole-like trapping [*29–31*], field redistribution, and topology-controlled wave manipulation. In addition to clarifying the role of anisotropy in shaping wormhole-like optical geometries, this mechanism also suggests practical functionalities including beam splitting, absorbing-like switching, and curvature-assisted focusing. This work points to a new way of using anisotropy in transformation optics to engineer tunable optical spaces with controllable topology, singularity structure, and associated wave functionalities, opening possibilities for topology-inspired photonic devices and singularity-enabled wave control.

## Conformal mapping with dual black holes

We begin by examining the optical space generated by a logarithmic conformal transformation, which gives rise to two singular points that act as black-hole–like absorbers for light [*32*]:

$$w(z) = z + \alpha(ln(z+\beta) - ln\,(z-\beta)). \tag{1}$$

where $w = u + iv$ represents the coordinates in virtual space, $z = x + iy$ represents the coordinates in physical space, $\alpha$ and $\beta$ are parameters of this transformation. This conformal transformation incorporates two logarithmic terms and a linear term. With finite values of $\alpha$ and $\beta$, this mapping ensures $w(z) = z$ at infinity, introducing two singularities in the physical space at locations $z = \beta$ and $z = -\beta$.

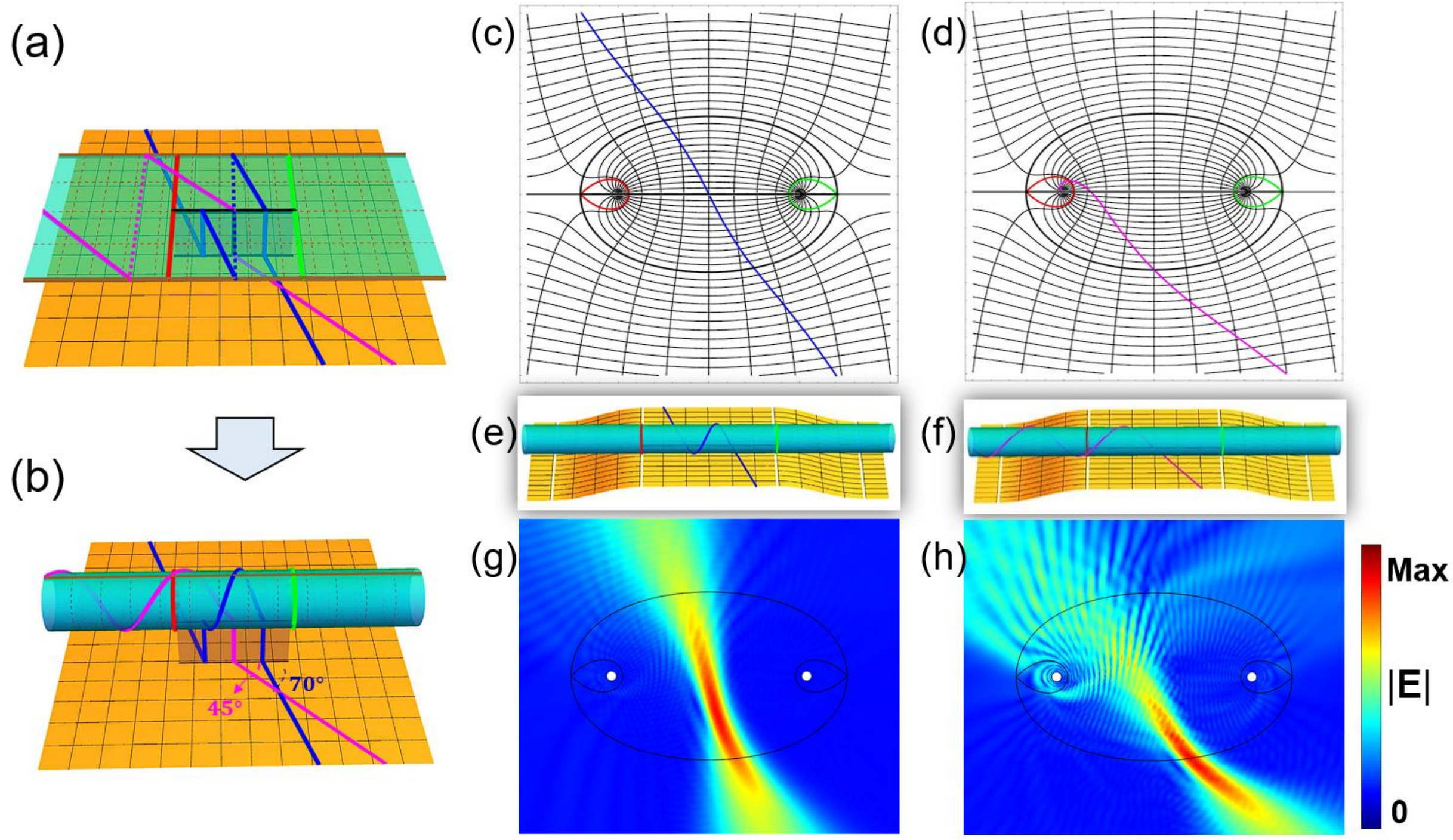

**Fig 1. The virtual and physical spaces of the conformal transformation with dual black holes.** (a) and (b) depict the two behaviors of light rays in the virtual space. The existence of periodic boundary conditions makes the upper Riemann sheet equivalent to a cylindrical surface. (c) and (d) illustrate the ray tracing of two scenarios in the physical space. (e) and (f) show the virtual space for $\alpha = 0.5$ and $\beta = 1$. (g) and (h) show simulations of a Gaussian beam with $\lambda = 0.1$ in the physical space. The black bold oval in the z plane correspond to the branch cut in (a) and (b). The interior of this circle relates to the upper Riemann sheet of the virtual space, while the exterior corresponds to the lower Riemann sheet.

As illustrated in Fig. 1(a), the virtual space consists of two Riemann sheets connected by a branch cut. The lower sheet is unbounded in both directions, whereas the upper sheet is finite in the transverse direction and periodic along its boundaries. This periodicity renders the upper Riemann sheet topologically equivalent to a cylinder, allowing light rays to circulate indefinitely once they enter this sheet [*33*]. In virtual space, light interacting with the branch cut exhibits two qualitatively distinct dynamical behaviors, depending on its incident direction. For certain small angles, rays cross the branch cut, temporarily access the cylindrical sheet, and subsequently return to the lower sheet, preserving global propagation, as shown by the blue lines in Fig. 1(a) and Fig. 1(b). In contrast, for small incident angles, rays are redirected toward the black hole boundaries (red and green circles) of the cylindrical sheet, where they propagate indefinitely and never return to the lower sheet as shown by the magenta lines in Fig. 1(a) and Fig. 1(b).

This virtual-space dynamics is directly inherited by physical space. Hamiltonian ray tracing [*34,35*] confirms that rays in physical space either complete a closed excursion on the upper Riemann sheet and re-emerge into the lower sheet, or are permanently captured by the singularities, faithfully reflecting the cylindrical topology of the virtual space (Figs. 1(c) and 1(d)). Full-wave simulations of transverse-electric (TE) Gaussian beams ($\lambda = 0.1$, $\alpha = 0.5$, $\beta = 1$) further confirm this behavior, showing that waves either traverse the branch cut or are predominantly captured by the singularities (Figs. 1(g) and 1(h)). Because a Gaussian beam possesses a finite angular spectrum, not all wave-vector components satisfy the trapping condition of the black-hole-like singularities. Consequently, in addition to the trapped components, some field components remain visible as scattered or escaping waves. Overall, this logarithmic conformal transformation naturally generates a pair of optical black holes, whose capture and transmission properties are consistently described by virtual-space topology, Hamiltonian ray dynamics, and full-wave simulations.

**Torus Projection**

To go beyond isolated optical black holes, we now examine how the topology of the virtual space can be altered to connect the two black hole endpoints in virtual space. The key idea is to map the cylindrical Riemann sheet, which originally has two independent asymptotic ends, onto a closed toroidal virtual geometry. This connection is achieved through a toroidal projection, inspired by the role of stereographic projection in Maxwell's fisheye lens [*36,37*]. While stereographic projection maps a closed spherical surface onto a plane, the torus projection maps a closed toroidal surface onto a cylindrical space, as shown in Fig. 2(a-d). In doing so, it introduces a geometric throat that links the two ends of the cylinder, which previously corresponded to independent black holes in virtual space.

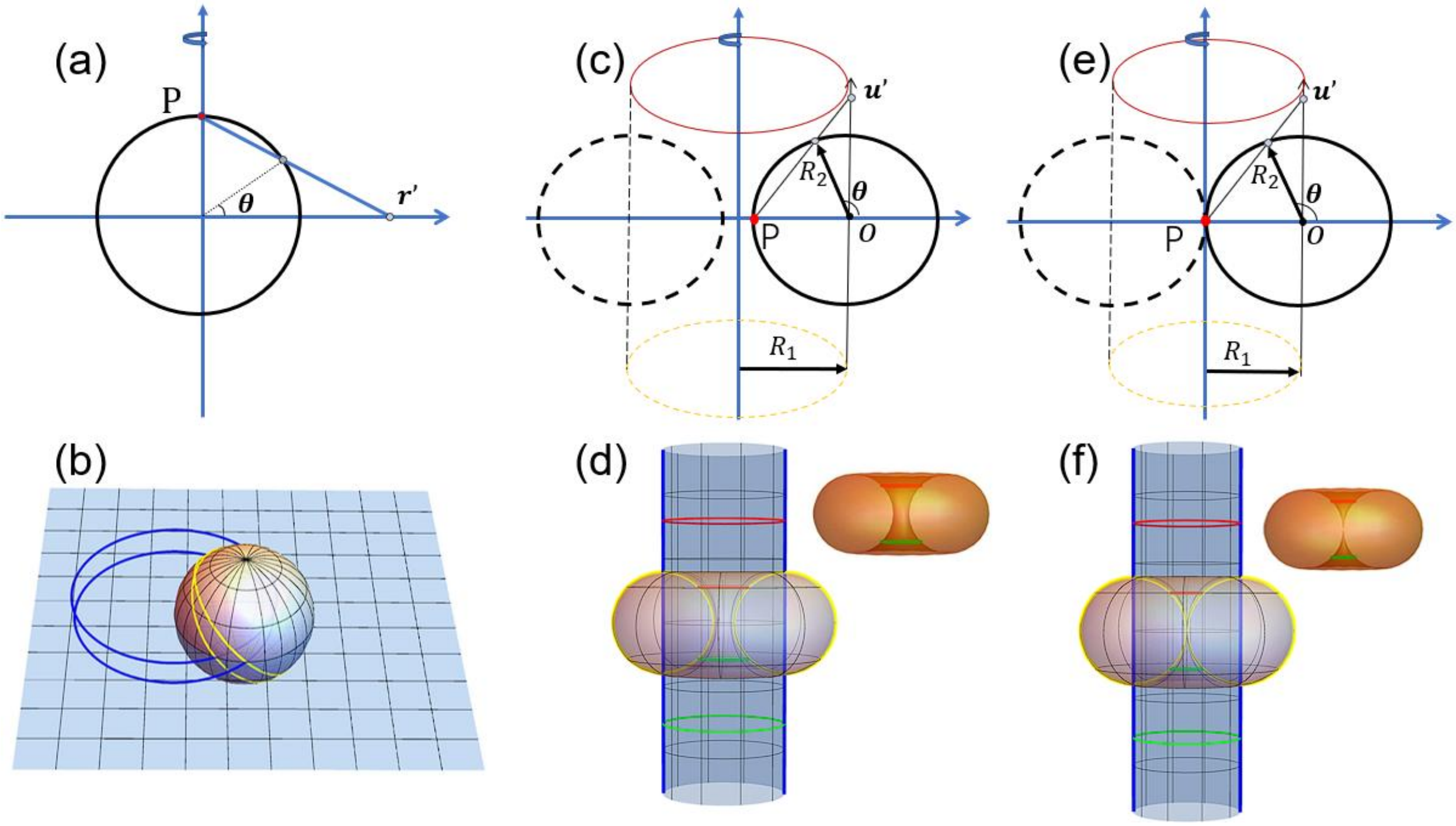

**Fig 2. stereographic and torus projections.** (a) Cross-sectional view of the stereographic projection. (b) Projection trajectories (blue lines) of geodesics on the circle (yellow lines) under spherical projection. (c) Cross-sectional view of the ring torus and (e)the horn torus projection. (d) and (f) show the projection trajectories on the cylindrical surface, respectively. The red and green circles represent the two black holes in the conformal mapping with dual black holes.

We now provide a detailed description of the torus projection. A torus is characterized by two radii: the major radius $R_1$ and the minor radius $R_2$. As illustrated in Fig. 2(c), a torus can be generated by rotating a circular cross-section of radius $R_2$ around a central axis located at a distance $R_1$, in a manner analogous to the formation of a sphere by rotating a circle. Similarly, we identify a projection point $P$ on this toroidal cross-section and project all points from the surface of the ring torus, with a major radius of $R_1$ and a minor radius of $R_2$, onto the cylindrical surface with a radius of $R_1$. In cylindrical coordinates $(u', \varphi')$, the mapping reads

$$\begin{cases} u' = R_2 \dfrac{\sin(\theta)}{\cos(\theta) + 1} \\ \varphi' = \varphi \end{cases} \tag{2}$$

where $u'$ represents the axial parameter of the cylindrical surface. The metric on the torus is given by $ds^2 = {R_2}^2 d\theta^2 + (R_1 + R_2 \cos(\theta))^2 d\varphi^2$, while the metric on the cylindrical surface with anisotropic refractive index distribution can be expressed as

$$ds'^2 = n_{u'}^2 du'^2 + n_{\varphi'}^2 {R_1}^2 d\varphi'^2 \tag{3}$$

We apply the torus projection mapping to the metric of the toroidal surface with anisotropic refractive index distribution:

$$ds'^2 = n_{u'}^2 R_2^2 \frac{1}{\left(1 + \cos(\theta)\right)^2} d\theta^2 + n_{\varphi'}^2 {R_1}^2 d\varphi^2 \tag{4}$$

To replicate the trajectory from a toroidal surface onto a cylindrical surface, we can obtain $n_{u'} = 1 + \cos(\theta) = 1 + \frac{R_2^2 - u'^2}{R_2^2 + u'^2}$ and $n_{\varphi'} = 1 + \frac{R_2}{R_1} \cdot \cos(\theta) = 1 + \frac{R_2}{R_1} \cdot \frac{R_2^2 - u'^2}{R_2^2 + u'^2}$.

From a geometrical standpoint, the torus projection does not merely reshape ray trajectories, but changes the global connectivity of the virtual space.

The upper Riemann sheet, which originally terminates at two singular endpoints, is mapped onto a closed toroidal virtual geometry, where geodesic trajectories can circulate along the toroidal surface. Thus, the two asymptotic ends are no longer independent terminations in virtual space, but are associated with a common closed manifold. This provides the geometrical basis for constructing a wormhole-like virtual optical space, whose opening and closure will be quantified below by the ratio between the torus parameters $R_2$ and $R_1$.

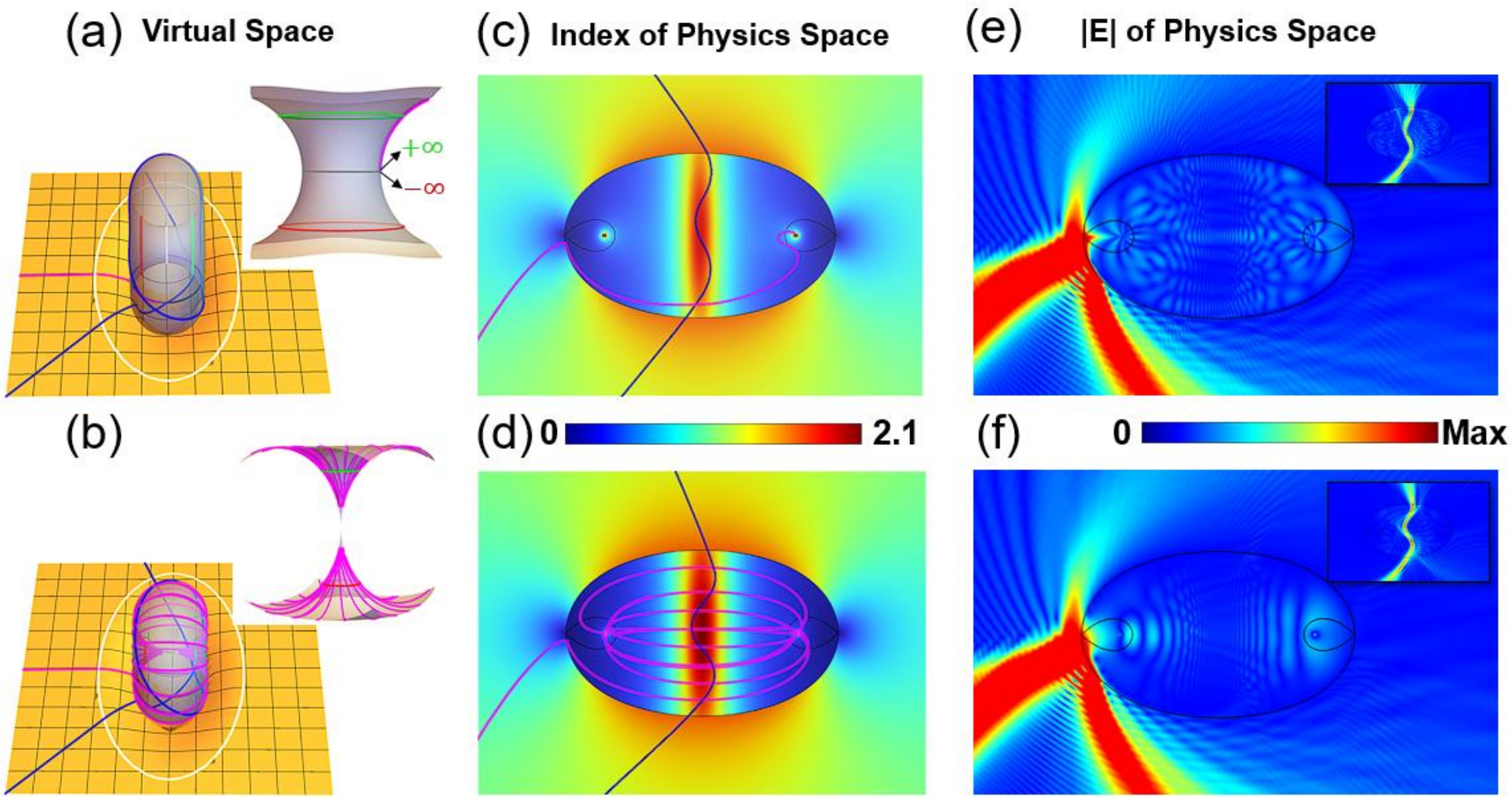

**Fig. 3. Construction of a torus-projected virtual wormhole and its electromagnetic signatures.** (a, b) Ray trajectories in virtual space for $c = \frac{2}{3}$ and $c = 1$, showing the open-throat and closed-

throat cases, respectively. (c, d) Corresponding ray trajectories and refractive-index distributions in physical space. (e, f) Full-wave simulations under Gaussian-beam excitation.

**Electromagnetic Wormhole**

We define the ratio $\frac{R_2}{R_1}$ as $c$, which controls the geometry of the torus and, consequently, the topology of the projected cylindrical space. Therefore, the refractive index on the cylinder in virtual space can be written as:

$$n_{u'} = 1 + \frac{c^2 {R_1}^2 - u'^2}{c^2 {R_1}^2 + u'^2},\ n_{\varphi'} = 1 + c \cdot \frac{c^2 {R_1}^2 - u'^2}{c^2 {R_1}^2 + u'^2} \tag{5}$$

where the anisotropy explicitly depends on the parameter c. When $c = 1$, the torus degenerates into a horn torus and the refractive index in virtual space becomes isotropic, satisfying $n_{u'} = n_{\varphi'}$. In this limit, radial and angular directions are equivalent up to a local scale factor, and the virtual-space geometry preserves conformal symmetry. Consequently, the transport channel collapses into an isolated singular point, and the two asymptotic ends of the cylindrical space remain disconnected. In contrast, when $c < 1$, the torus projection introduces an intrinsic anisotropy into the virtual-space metric, leading to $n_{u'} \neq n_{\varphi'}$. This anisotropy breaks the conformal symmetry of the virtual space: radial and angular directions are no longer equivalent, and scale–rotation covariance is lost. As a result, the two asymptotic ends of the virtual cylindrical space become connected through a toroidal manifold, giving rise to a wormhole-like topology in virtual space. Therefore, by tuning the parameter c, the wormhole can be continuously opened or closed in virtual space, realizing a topological transition between a horn torus with disconnected asymptotic regions and a ring torus with a connected throat.

In our case, $u'$ is the axial coordinate of the cylinder, corresponding to the $u$-direction in the conformal transformation with dual black holes. To ensure consistency with the conformal mapping, the major radius $R_1$ of the torus is identified with the mapping parameter $\alpha$. Thus, the refractive index for the cylinder can be expressed as:

$$n_{u'} = 1 + \frac{c^2 \alpha^2 - u^2}{c^2 \alpha^2 + u^2}, \qquad n_{\varphi'} = 1 + c \cdot \frac{c^2 \alpha^2 - u^2}{c^2 \alpha^2 + u^2} \tag{6}$$

The corresponding refractive index distribution in physical space is more complex, and the detailed derivation and expression can be found in Appendix A.

The effect of this topology control is first illustrated in virtual space, as shown in Figs. 3(a) and 3(b). For $c = 2/3$, the torus projection opens a virtual throat between the two asymptotic ends of the cylindrical space. Rays entering the interior of the ring are then driven toward the asymptotic region in the cylindrical representation. Although the trajectory remains continuous on the toroidal surface, its cylindrical representation undergoes a transition from $u' = +\infty$ to $u' = -\infty$ as the ray passes through the inner throat region. Therefore, the two infinities of the cylindrical representation are connected only through the toroidal geometry. On the cylinder, this process cannot be shown as an ordinary finite and continuous trajectory; instead, the ray is projected toward an infinity point, which corresponds to a black-hole-like singularity and produces the trapping effect near this region, as shown in Fig. 3(c). A detailed geometrical analysis of this infinity-to-infinity transition is provided in Appendix B. For $c = 1$, the torus degenerates into a horn torus, the virtual throat collapses, and the infinity-to-infinity transition disappears, indicating that the two asymptotic regions remain disconnected. As a result, rays can no longer evolve continuously from one end to the other, but are instead turned back near the outer surface of the torus, leading to a back-and-forth propagation along the outer region. The corresponding ray trajectories and refractive-index distributions in physical space are shown in Fig. 3(d). The full-wave simulations under Gaussian-beam excitation, shown in Figs. 3(e) and 3(f), further support this interpretation. For $c < 1$, the field exhibits pronounced localization and petal-shaped intensity patterns near the singularities, indicating a trapping response induced by the torus projection. When $c = 1$, the field distribution becomes more regular and point-like, consistent with the closure of the virtual throat and the suppression of the topology-induced trapping response.

In addition to the black-hole-like trapping effect, the blue trajectories in Figs. 3(c) and 3(d) indicate that both topologies still preserve an intrinsic feature of the outer toroidal surface. These rays are not driven toward the projected infinity; instead, they remain confined near the outer surface, are bent back by the toroidal curvature, and eventually return to the lower Riemann sheet to continue propagating. The corresponding wave behavior is likewise visible in the small panels of Figs. 3(e) and 3(f). Meanwhile, the outer toroidal surface retains an important self-focusing property [*34*], which suggests a possible focusing functionality of the torus-projected geometry and may be useful for beam concentration. A more detailed discussion is given in Appendix C.

To further examine the sensitivity of this black-hole-like response to anisotropy, we consider the field patterns for $c = 1, 0.98$, and 0.8, shown in Fig. 4. The results indicate that the black-hole-like effect emerges very sensitively near $c = 1$: even a slight deviation from the isotropic limit, such as from $c = 1$ to $c = 0.98$, already produces a pronounced change in the field structure around the singularities. As $c$ decreases further, this angular modulation becomes increasingly prominent, indicating the rapid onset and strengthening of the singularity-induced black-hole-like response.

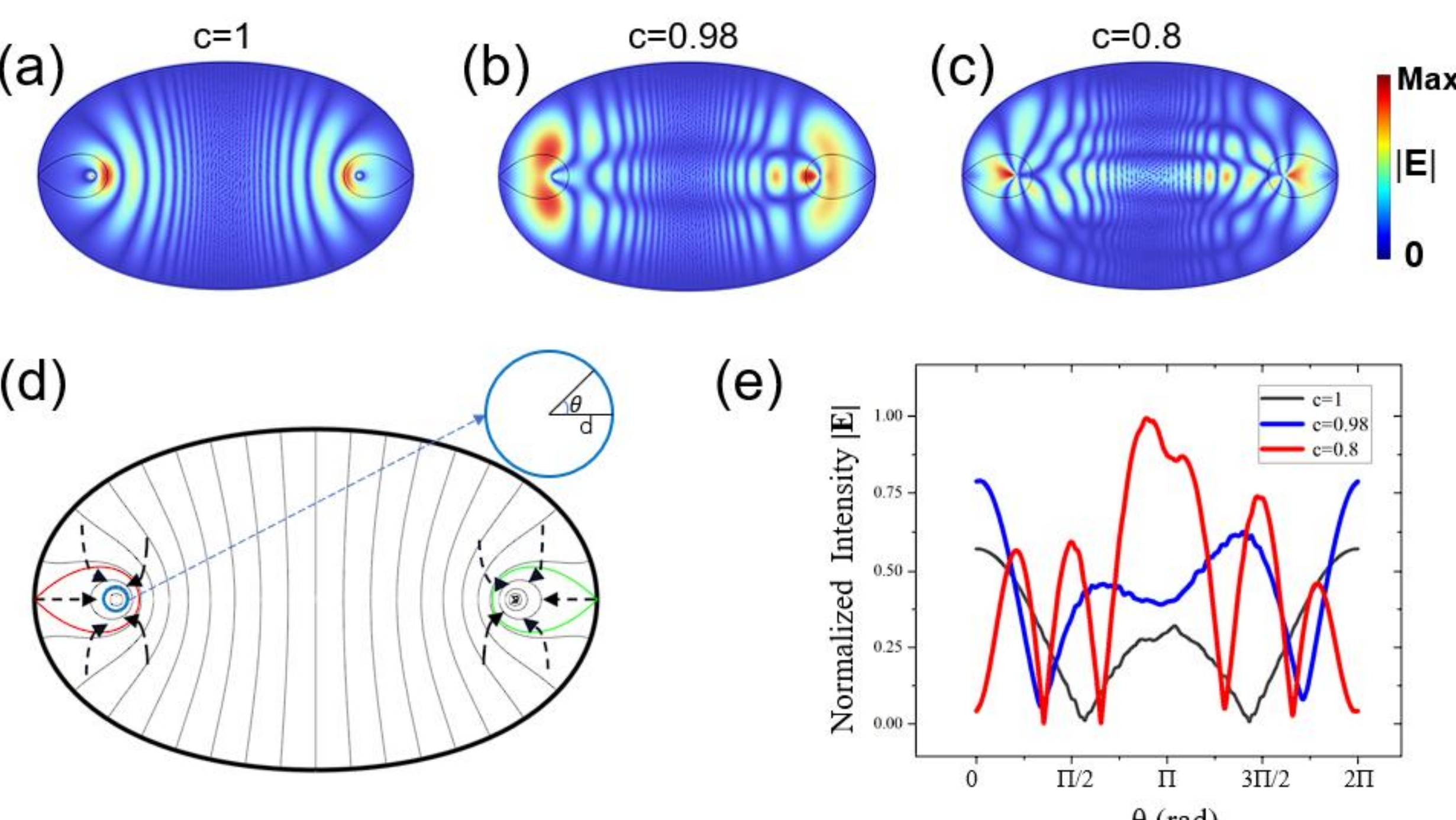

**Fig. 4. Evolution of eigenmode field patterns and singularity reconstruction induced by anisotropy.** (a–c) Electric-field amplitude distributions of representative eigenmodes for c=1, 0.98,

and 0.8, obtained from eigenfrequency analysis. (d) Schematic illustration of the field-pattern evolution as $c$ decreases. (e) Normalized electric-field intensity |E| sampled along a circular path (radius d = 0.1) around one singularity. As $c$ decreases, the enhanced |E| near the singularities indicates a stronger singularity-induced confinement and black-hole-like trapping response

**Switching Beam splitter enabled by anisotropic singularity reconstruction**

Based on the black-hole-like trapping effect induced by the topology transition in virtual space, we further consider using the anisotropic singularity reconstruction as a switching mechanism to control electromagnetic-wave propagation in physical space. Here we demonstrate that, by tuning the topology of the toroidal virtual surface, the same structure can be switched between a beam-splitting functionality and an absorbing functionality. In the previous configuration, the refractive-index contrast between the lower and upper Riemann sheets at the two ends of the branch cut is too large, so that a Gaussian beam incident from left to right is mostly reflected. To facilitate coupling of the wave into the inner toroidal surface, we instead adopt another logarithmic conformal transformation,

$$w(z) = z + \alpha(ln(z-\beta) - ln(z+\beta)). \quad (7)$$

The only difference from Eq. (1) is that the orientation of the branch cut is rotated by 90 degrees, while the cylindrical direction of the upper Riemann sheet remains unchanged [32]. As a result, electromagnetic waves can enter the inner toroidal surface more efficiently. The corresponding refractive-index distribution can be derived in a manner analogous to that described in Appendix A. As mentioned above, the parameter $\alpha$ should remain consistent with the major radius $R_1$ of the torus. Therefore, we again tune the topology by varying the minor radius $R_2$, namely by changing the ratio $c$, which controls the toroidal geometry and hence the topology of the virtual space. As shown in Figs. 5(a)–5(c), we choose $c = 1.25$, 1.00, and 0.75 to represent the closed, threshold, and open states, respectively.

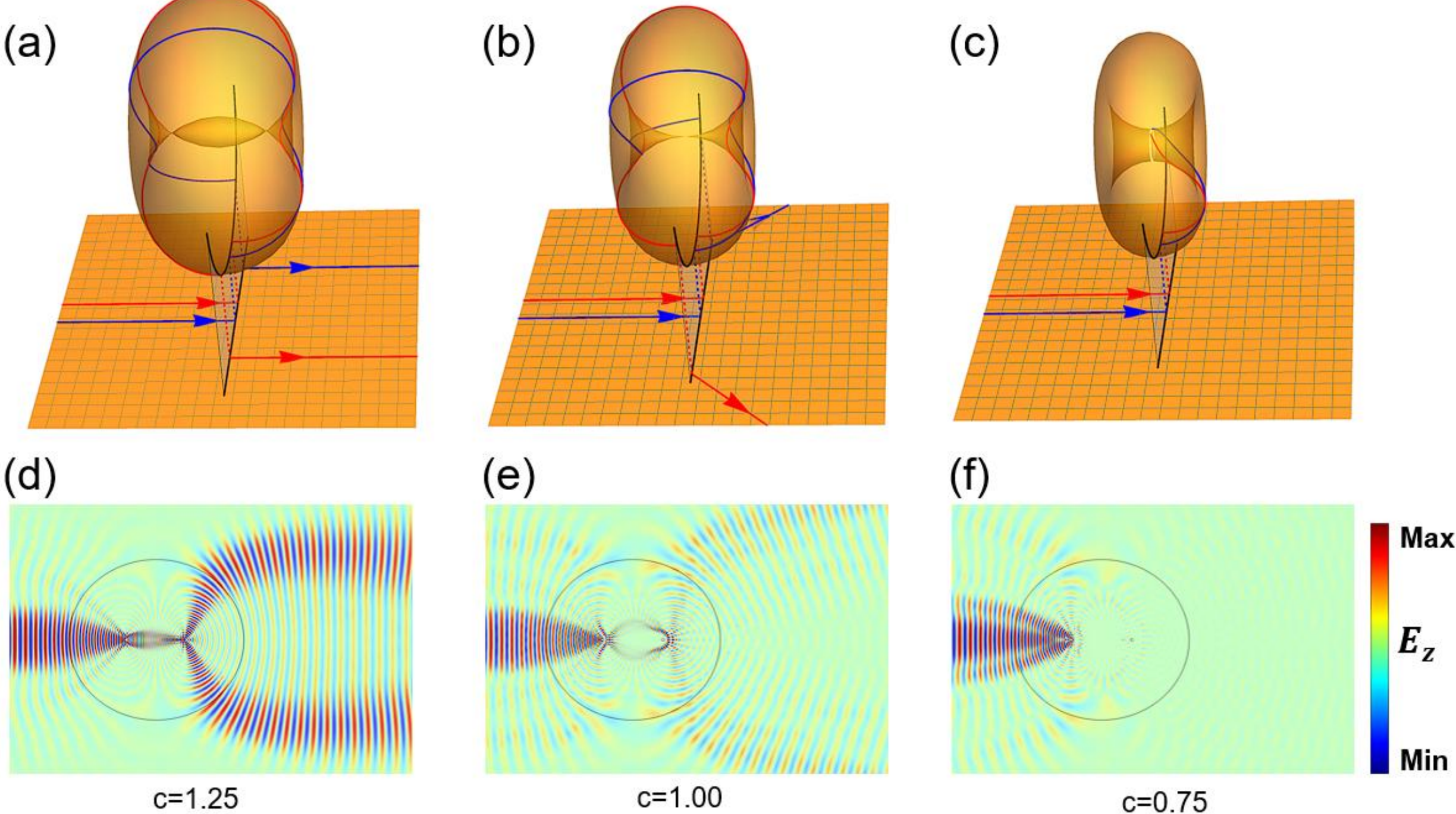

**Fig. 5. Beam-splitting and absorbing-like switching enabled by Anisotropic Singularity Reconstruction.** (a–c) Ray-trajectory switching from beam splitting to trapping/absorbing-like

behavior for $c = 1.25, 1.00$, and $0.75$, respectively. (d–f) Corresponding full-wave simulations for the three states under Gaussian-beam excitation (($\lambda = 0.5$, $\alpha = 4$, $\beta = 1$).

When $c = 1.25$, the incident rays enter the upper Riemann sheet from the lower sheet and then propagates along the toroidal surface. Because the toroidal surface guides geodesics differently for rays entering at different positions, while still preserving an overall symmetry, the injected beam is separated into two distinct ray families, represented by the blue and red trajectories in Fig. 5(a). These two rays eventually return to the lower Riemann sheet with approximately the same overall propagation direction, but as two spatially separated outgoing beams, thereby realizing a beam-splitting effect. When $c = 1$, the topology has not yet changed and the black-hole-like trapping effect is still absent. Therefore, the beam-splitting behavior persists. However, because the geodesic trajectories on the torus depend sensitively on the ratio $c$, the two outgoing beams returning to the lower Riemann sheet now emerge with a much wider splitting angle, as shown in Fig. 5(b). Finally, when $c = 0.75$, the topology transition of the virtual space takes place. The geodesics on the toroidal surface now exhibit an infinity-to-infinity transition, so that the rays are driven toward the projected infinity, which manifests in physical space as a black-hole-like trapping effect near the singularities, and can no longer return to the lower Riemann sheet. As a result, the beam-splitting functionality is switched off and replaced by an absorbing-like response, as shown in Fig. 5(c). We also perform full-wave simulations using a Gaussian beam with wavelength $\lambda$=0.5. The results, shown in Figs. 5(d)–5(f), confirm that the black-hole-like trapping effect induced by the topology transition in virtual space indeed provides a switching mechanism between the beam-splitting and absorbing states.

**Conclusion**

In this work, we have proposed an electromagnetic wormhole based on a torus projection within the framework of transformation optics, in which two black-hole–like singularities remain spatially separated in physical space while becoming topologically connected in virtual space. Beyond the construction of a specific wormhole geometry, our results reveal a geometric mechanism by which anisotropy reconstructs optical singularities and induces a topological transition of the virtual space. By tuning a single parameter $c$, the virtual-space metric evolves from an isotropic, conformally symmetric horn-torus geometry to an intrinsically anisotropic ring-torus geometry. This transition breaks the equivalence between radial and angular directions, converts point-like singular terminations into extended transport channels, and changes the global connectivity of the virtual cylindrical space. Building on this mechanism, we further demonstrated that anisotropic singularity reconstruction can be used as a switching principle for wave functionalities in physical space. In particular, by tuning the toroidal topology, the same structure can be switched between a beam-splitting state and an absorbing-like state, providing a concrete example of topology-controlled wave manipulation enabled by the virtual-space transition. These results show that anisotropy in transformation optics is not merely a quantitative modification of refractive index, but a geometric degree of freedom for controlling singularity structure, virtual-space topology, and their observable electromagnetic consequences in physical space. More broadly, this work points to a route toward tunable topology-inspired photonic platforms in which singularity engineering, wave trapping, beam switching, and curvature-assisted functionalities can be integrated within a unified framework.

**Appendix A. Refractive index of the wormhole in physical space**

When integrating the conformal transformation with the torus projection to form a wormhole, additional complexities must be addressed. Firstly, parameters $\alpha$ in the conformal transformation govern the period in the upper Riemann sheet, which corresponds to the radius of the cylindrical surface. The main radius $R_1$ of the projected torus must be equal to $\alpha$ in conformal transformation. Secondly, the calculation of the refractive index distribution in conformal transformations, given by $n = n_w|\frac{dw}{dz}|$, satisfies the Cauchy-Riemann conditions if and only if the refractive index distribution $n_w$ in the virtual space is isotropic. If $n_w$ is anisotropic in the virtual space, this formula is no longer applicable, and a general coordinate transformation is required to compute the corresponding refractive index distribution in the physical space.

To begin with, we examine the simplified case of a horn wormhole, where *c*=1, yielding $n_{u'} = n_{\varphi'} = \frac{2\alpha^2}{\alpha^2+u^2}$. Due to the isotropic material distribution in the virtual space, we can straightforwardly apply the formula $n = n_w|\frac{dw}{dz}|$ to obtain an isotropic refractive index distribution in the physical space:

$$\mathrm{n}_{in} = \frac{2\alpha^2}{\alpha^2 + u^2} \cdot \left| \frac{\beta^2 + 2\alpha\beta - z^2}{\beta^2 - z^2} \right| \qquad (A1)$$

where $z = x + iy$ and $u = x + \frac{\alpha}{2} * \ln\left(\frac{(x+1)^2+y^2}{(x-1)^2+y^2}\right)$ as determined by the conformal transformation. The refractive index distribution outside the wormhole is given by:

$$\mathrm{n}_{out} = \left| \frac{\beta^2 + 2\alpha\beta - z^2}{\beta^2 - z^2} \right| = \frac{\sqrt{(\beta^2 + 2\alpha\beta - x^2 + y^2)^2 + 4x^2y^2}}{\sqrt{(\beta^2 - x^2 + y^2)^2 + 4x^2y^2}} \qquad (A2).$$

In the main text, we set $\alpha$=0.5 and β=1 to align with the previous dual black hole simulation. The refractive index of a ring wormhole is not isotropic, making the calculations more complex. Firstly, in virtual space, the line element of a cylindrical surface is:

$$ds'^2 = n_{r'}^2 dr'^2 + n_{\varphi'}^2 {R_1}^2 d\varphi'^2 \qquad (A3)$$

Given $du' = du$ and $R_1 d\varphi = dv$, the line element in virtual space can also be expressed in terms of $u$ and $v$ as:

$$ds'^2 = n_{u'}^2 du^2 + n_{\varphi'}^2 dv^2 \qquad (A4)$$

However, to achieve a material realization in physical space that replicates the same form of the line element as in virtual space, we need to determine the metric tensor form in physical space and subsequently derive the corresponding electromagnetic parameters. In physical space, the line element is expressed as:

$$ds'^2 = n_{uv} dx^u dx^v \qquad (A5)$$

The transformation between the coordinates of physical and virtual space is determined by the conformal transformation equation:

$$\begin{cases} \mathrm{u} = x + \frac{\alpha}{2} \cdot \ln\left(\frac{(x + \beta)^2 + y^2}{(x - \beta)^2 + y^2}\right) \\ v = y + \alpha \cdot \left(\mathrm{Pi} + \mathrm{Tan}^{-1}\left(\frac{y}{x + \beta}\right) - \mathrm{Tan}^{-1}\left(\frac{y}{x - \beta}\right)\right) \end{cases} \qquad (A6)$$

From simple coordinate substitutions, we know:

$$ds'^2 = \left(n_{u'}^2 \cdot \left(\frac{\partial u}{\partial x}\right)^2 + n_{\varphi'}^2 \cdot \left(\frac{\partial v}{\partial x}\right)^2\right) dx^2 + 2\left(n_{u'}^2 \cdot \left(\frac{\partial u}{\partial x}\right)\left(\frac{\partial u}{\partial y}\right) + n_{\varphi'}^2 \cdot \left(\frac{\partial v}{\partial x}\right)\left(\frac{\partial v}{\partial y}\right)\right) dxdy + \left(n_{u'}^2 \cdot \left(\frac{\partial u}{\partial y}\right)^2 + n_{\varphi'}^2 \cdot \left(\frac{\partial v}{\partial y}\right)^2\right) dy^2$$

By substituting Eq. (A6) into the above expression, we can obtain the refractive index distribution in physical space.

$$n = \begin{bmatrix} n_{xx} & n_{xy} & \\ n_{xy} & n_{yy} & \\ & & 1 \end{bmatrix} \tag{A7}$$

$$n_{xx} = \frac{16x^2y^2\alpha^2\left((-1+c)u^2 - c^2(1+c)\alpha^2\right)^2\beta^2 + 4c^4\alpha^4\left(x^4 + 2x^2\left(y^2 - \beta(\alpha+\beta)\right) + (y^2+\beta^2)\left(y^2 + \beta(2\alpha+\beta)\right)\right)^2}{(u^2 + c^2\alpha^2)^2(y^2 + (x-\beta)^2)^2(y^2 + (x+\beta)^2)^2}$$

$$n_{xy} = \frac{2(-1+c)xy\alpha\left((-1+c)u^4 - 2c^2(1+c)u^2\alpha^2 + c^4(3+c)\alpha^4\right)\beta\left(x^4 + 2x^2\left(y^2 - \beta(\alpha+\beta)\right) + (y^2+\beta^2)\left(y^2 + \beta(2\alpha+\beta)\right)\right)}{(u^2 + c^2\alpha^2)^2(x^2 + y^2 - 2x\beta + \beta^2)^2(x^2 + y^2 + 2x\beta + \beta^2)^2}$$

$$n_{yy} = \frac{64c^4x^2y^2\alpha^6\beta^2 + \left((-1+c)u^2 - c^2(1+c)\alpha^2\right)^2\left(x^4 + 2x^2\left(y^2 - \beta(\alpha+\beta)\right) + (y^2+\beta^2)\left(y^2 + \beta(2\alpha+\beta)\right)\right)^2}{(u^2 + c^2\alpha^2)^2(y^2 + (x-\beta)^2)^2(y^2 + (x+\beta)^2)^2}$$

where $u$ and $v$ are given by Eq. (A6). When $c = 1$, through calculations, we find that $n_{xx} = n_{yy}$, $n_{xy} = 0$ and the refractive index returns to the isotropic result obtained through conformal mapping.

## Appendix B. Geometric Ray Trajectories in Open-Ring Wormholes: A Transition Between Infinities

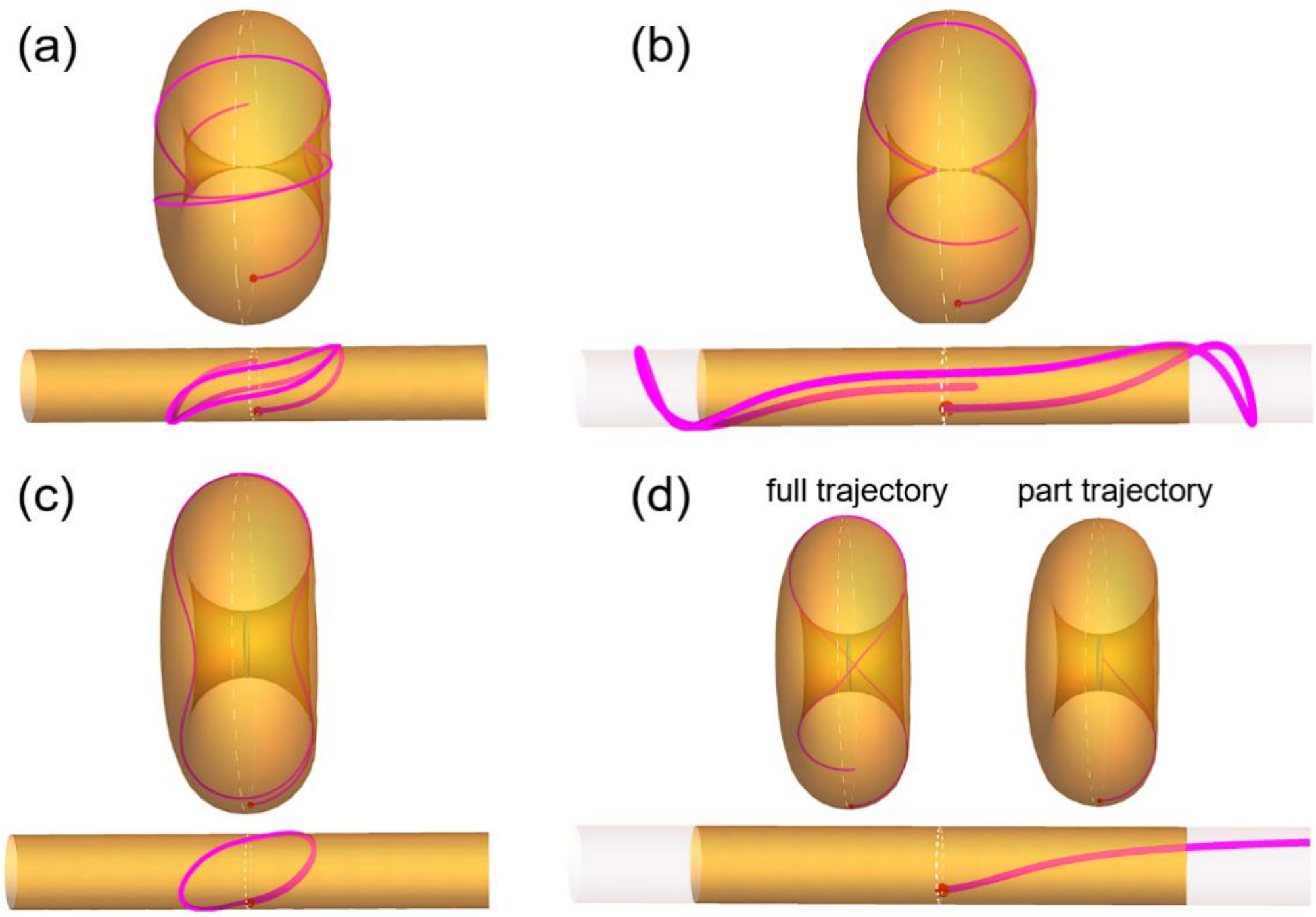

**Fig. 6. Projection of the torus surface trajectory onto the cylindrical surface.** (a) Ray trajectory at small angle incidence; (b) Ray trajectory at large angle incidence for the singularity of the horn-shaped torus. (c) Ray trajectory at small angle incidence; (d) Ray trajectory at large angle incidence for the throat (gray circle) of the ring torus. The red dots indicate the ray emission points.

This appendix explains how the infinity-to-infinity transition in an open-ring wormhole arises from the torus projection, and how this transition is represented as black-hole-like trapping after projection into physical space. When the light does not pass through infinity (i.e., the ends of the cylinder), as illustrated in Figs.6 (a-c), the light follows a curved path due to the surface curvature of the ring, continuously circling around the ring. Through the torus projection described in the main text, we can project the complete trajectory of the ray on the ring to the cylindrical surface by filling the refractive index, as discussed in Section 3. This projection is shown on the cylindrical surface in the lower part of Figs. 6 (a-c). (To simplify our analysis, the light source is positioned at virtual-space coordinates (0,0)).

When $c = 1$, corresponding to a horn-shaped ring, due to the inherent structural properties of the ring's surface, no ray will pass through the singularity. Even rays that are incident directly towards the ring singularity will reverse their direction just before reaching the singularity, as shown in Fig. 6(b). On the corresponding cylindrical diagram, the rays also turn back at a distant point on the cylinder. However, when $c = 2/3$, and the rays do not pass through the interior of the ring, as shown in Fig. 6(c), the complete trajectory can still be represented correspondingly on the ring surface with the refractive index material. Once the rays pass through the interior of the ring, as shown in Fig. 6(d), the cylindrical surface fails to accurately reflect the complete trajectory of the rays within the ring. Instead, the rays are projected towards infinity. If we project rays near this infinity point back onto the ring surface, they will converge on an infinitesimally small circle centered at the gray region. This circle corresponds to the singularity where the horn-shaped wormhole is expanded.

We project the remaining ray trajectories from Fig. 7(d) back onto the cylindrical surface, as shown in Fig. 7. As seen in Fig. 7(a), the red ray on the cylinder propagates to the right end; however, in Fig. 7(b), the ray propagates back from the left end of the cylinder. This indicates that the ray undergoes a transition at end, moving from one infinity to another. Furthermore, from Figs. 7 (b, c), the ray does not encounter the previous singularity, ensuring that the light remains continuous. From Figs. 7s (c, d), the ray again passes through the singularity where the ring expands, transitioning from one infinity to another. The complete trajectory on the ring, as well as the trajectory on the cylindrical surface with the material, is shown in Fig. 7(e).

This transition from one infinity to another is actually caused by the torus projection itself. As mentioned in the main text:

$$u' = R_2 \frac{\sin(\theta)}{\cos(\theta) + 1} \tag{B1}$$

By applying a simple half-angle transformation, we get:

$$u' = R_2 * \tan\left(\frac{\theta}{2}\right) \tag{B2}$$

When $\theta$ transitions from $\theta \to \pi^+$ to $\theta \to \pi^-$(The ray passes through the gray circle), we observe that $u'$ moves from $+\infty$ to $-\infty$. Thus, the torus projection naturally produces an infinity-to-infinity transition in the cylindrical virtual representation. In physical space, this transition is not observed as an ordinary finite trajectory connecting two points. Instead, it is projected as ray accumulation near the singularities, giving rise to the apparent black-hole-like trapping behavior discussed in the main text.

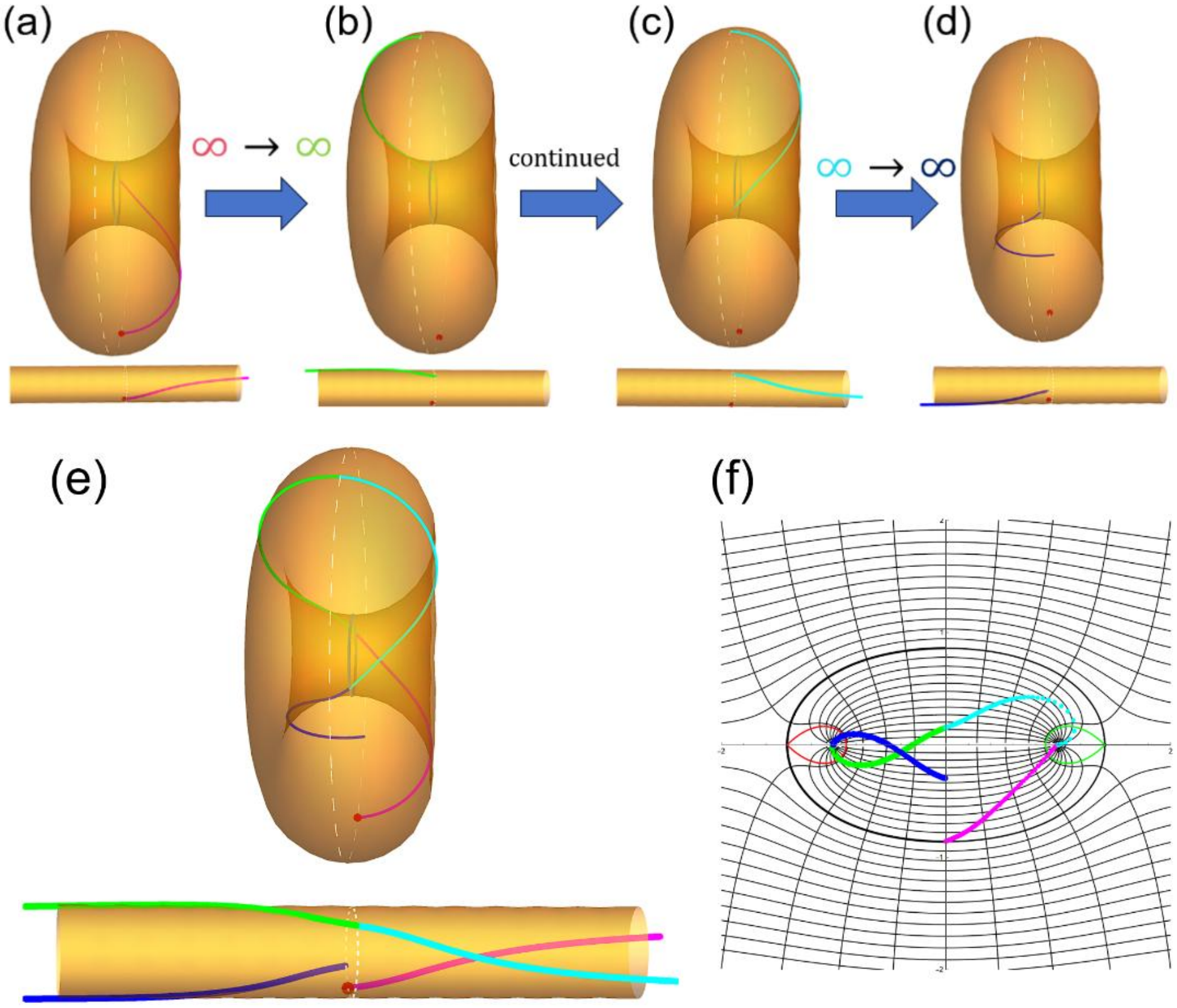

**Fig. 7 Projection of the torus surface trajectory onto the cylindrical surface**. Part trajectory of the ray for $\frac{\theta}{2}$ from (a) 0 to $\frac{\pi}{2}$ (b) $\frac{\pi}{2}$ to $\pi$ （c） $\pi$ to $\frac{3\pi}{2}$ (d) $\frac{3\pi}{2}$ to $2\pi$ (e) Complete projected trajectory of the ray for $\frac{\theta}{2}$ from 0 to $2\pi$. (f) Genetic Algorithm numerically reconstructed physical space trajectory.

To visualize the corresponding trajectory in physical space, we can map the trajectory from virtual space to physical space by inversely solving for the spatial coordinates (x, y) through Eq. (A6). However, this coordinate inversion entails solving a nonlinear system. To address this, we numerically determine the solutions at discrete points, reconstructing the trajectory. As shown in Fig. 7(f), the reconstructed physical-space trajectory exhibits an apparent discontinuous jump near the singularity. This discontinuity does not represent a physically continuous propagation path; rather, it indicates that the infinity-to-infinity transition in the toroidal virtual geometry cannot be faithfully represented as a finite trajectory in physical space. Consequently, the ray is effectively projected toward the singularity, giving rise to the black-hole-like trapping behavior discussed in the main text.

## Appendix C. Toroidal focusing effect in physical space

As noted in the main text, besides the black-hole-like trapping effect near the singularities, the torus-projected geometry also preserves a focusing feature inherited from the outer toroidal surface. This effect originates from the intrinsic curvature of the torus rather than from the infinity-to-infinity transition. Rays propagating near the outer toroidal surface are bent by the toroidal curvature and are refocused after a finite propagation distance, which appears in wave simulations as a sequence of intensity maxima.

This focusing behavior depends strongly on the torus aspect ratio $c = \frac{R_2}{R_1}$. Following Ref. [43], the angular focusing length on the toroidal surface can be rewritten in this notation as

$$\phi_f(c) = \pi\sqrt{\frac{c}{1+c}} \quad (C1)$$

and the corresponding number of focusing nodes in one round trip becomes

$$N(c) = \frac{2\pi}{\phi_f(c)} = 2\sqrt{1+\frac{1}{c}} \quad (C2)$$

These relations show that the number of focusing events increases as $c$ decreases, i.e., a thinner torus supports denser refocusing along the outer toroidal surface.

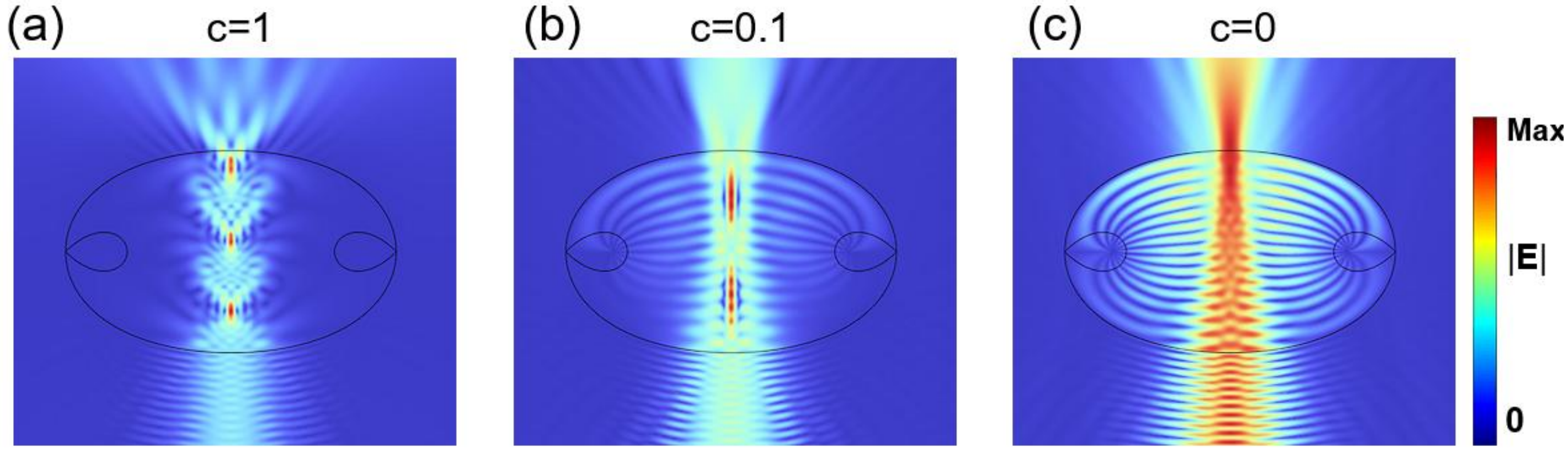

**Fig. 8. Evolution of curvature-assisted focusing with the torus aspect ratio $c$.** (a–c) Full-wave field distributions for $c = 1$, $0.1$, and $0$, respectively. Reducing $c$ increases the density of refocusing events on the outer toroidal surface, driving the field pattern from discrete focal spots toward a quasi-continuous focusing channel. In the limiting case $c = 0$, the pattern is dominated by line-like enhancement associated with the $n_{u'} = 0$; $n_{\varphi'} = 1$ limit.

Representative full-wave results are shown in Fig. 8. For $c = 1$, the field exhibits a discrete set of focal spots. When $c$ is reduced to 0.1, the number of focusing events increases significantly, so that the focal spots become densely packed and merge into a quasi-continuous focal line. In the limiting case $c = 0$, the effective refractive-index components become $n_{u'} = 0$ and $n_{\varphi'} = 1$. The system then approaches a zero-index-like response along the y-direction, and the field pattern is dominated by an extended line-like enhancement rather than isolated focal points. Therefore, as $c$ decreases, the toroidal focusing effect evolves from discrete multipoint focusing to focal-line formation, revealing another tunable wave-control property of the torus-projected geometry.

**Acknowledgments:** This work was supported by National Key Research and Development Program of China (Grant Nos. 2023YFA1407100 and 2020YFA0710100), National Natural Science Foundation of China (12404371), Jiangxi Provincial Natural Science Foundation (Grant No. 20224ACB201005). Wen Xiao also acknowledges the support of China Postdoctoral Science Foundation (GZC20240906). Basic and Applied Basic Research Foundation of Guangdong Province (2024A1515011724)

**Funding:**

National Key Research and Development Program of China 2023YFA1407100

National Key Research and Development Program of China 2020YFA0710100

National Natural Science Foundation of China 12404371

Jiangxi Provincial Natural Science Foundation 20224ACB201005

Basic and Applied Basic Research Foundation of Guangdong Province 2024A1515011724

**Competing interests:** Authors declare that they have no competing interests.

**Data and materials availability:** All data are available in the main text or the supplementary materials.